\documentclass[conference]{IEEEtran}
\IEEEoverridecommandlockouts
\usepackage{cite}
\usepackage{amsmath,amssymb,amsfonts}
\usepackage{graphicx}
\usepackage{textcomp}
\usepackage{xcolor}
\usepackage[dvipsnames]{xcolor}
\usepackage{enumitem}
\usepackage[caption=false,font=footnotesize,labelfont=sf,textfont=sf]{subfig}
\usepackage{amsmath,amsfonts}
\usepackage{graphicx}
\usepackage{textcomp}
\usepackage{threeparttable}
\usepackage{xcolor}
\usepackage{multirow}
\usepackage{pifont}
\usepackage{url}
\usepackage{bbding}
\usepackage{tikz}

\usepackage{listings}
\usepackage{xcolor}
\usepackage{framed}
\newcommand{\cmark}{\ding{51}}%
\newcommand{\xmark}{\ding{55}}%
\usepackage[table,xcdraw]{xcolor} 
\usepackage{colortbl}
\usepackage{orcidlink}
\usepackage[linesnumbered,ruled,vlined]{algorithm2e}
\usepackage[noend]{algpseudocode}
\newcommand{\bstars}{\textsuperscript{$\bigstar$}}

\newcommand{\blzs}{\textsuperscript{$\blacklozenge$}}
\newcommand{\btrs}{\textsuperscript{$\blacktriangle$}}

\newcommand{\nlbs}{\textsuperscript{$\nabla$}}

\usepackage[strict]{changepage}

\usepackage{framed}
\newcommand{\profuzz}{\texttt{\textbf{PROFUZZ}}}

\usepackage[most]{tcolorbox}

\def\BibTeX{{\rm B\kern-.05em{\sc i\kern-.025em b}\kern-.08em
    T\kern-.1667em\lower.7ex\hbox{E}\kern-.125emX}}

\usepackage{fancyhdr}
\fancypagestyle{firstpage}{

\fancyfoot[C]{\large{\textcolor{blue}{This paper has been accepted for publication at ICCAD 2025. This is an initial author version.\\ The final version is available at IEEE Xplore: https://doi.org/10.1109/ICCAD66269.2025.11240782}}}

}

\begin{document}

\title{Intelligent Graybox Fuzzing via ATPG-Guided Seed Generation and Submodule Analysis 
}

\author{
\IEEEauthorblockN{Raghul Saravanan\IEEEauthorrefmark{1}\orcidlink{0000-0002-8296-2144}, Sudipta Paria\IEEEauthorrefmark{2}\orcidlink{0009-0002-7726-8032}, Aritra Dasgupta\IEEEauthorrefmark{2}\orcidlink{0000-0002-0786-9185},
Swarup Bhunia\IEEEauthorrefmark{2}\orcidlink{0000-0001-6082-6961}, and Sai Manoj P D\IEEEauthorrefmark{1}\orcidlink{0000-0002-4417-2387}
\IEEEauthorblockA{\IEEEauthorrefmark{1}Department of Electrical and Computer Engineering, George Mason University, Fairfax, VA 22030, USA\\
\IEEEauthorrefmark{2}Department of Electrical and Computer Engineering, University of Florida, Gainesville, FL 32611, USA\\
\{rsaravan@gmu.edu, sudiptaparia@ufl.edu, aritradasgupta@ufl.edu, swarup@ece.ufl.edu, spudukot@gmu.edu\}} 
}
}

\maketitle
\thispagestyle{firstpage}

\begin{abstract}
Hardware Fuzzing emerged as one of the crucial techniques for finding security flaws in modern hardware designs by testing a wide range of input scenarios. One of the main challenges is creating high-quality input seeds that maximize coverage and speed up verification. Coverage-Guided Fuzzing (CGF) methods help explore designs more effectively, but they struggle to focus on specific parts of the hardware. Existing Directed Gray-box Fuzzing (DGF) techniques like DirectFuzz try to solve this by generating targeted tests, but it has major drawbacks, such as supporting only limited hardware description languages, not scaling well to large circuits, and having issues with abstraction mismatches. To address these problems, we introduce a novel framework \profuzz~that follows the DGF approach and combines fuzzing with Automatic Test Pattern Generation (ATPG) for more efficient fuzzing. By leveraging ATPG’s structural analysis capabilities, \profuzz~can generate precise input seeds that target specific design regions more effectively while maintaining high fuzzing throughput. Our experiments show that \profuzz~scales 30× better than DirectFuzz when handling multiple target sites, improves coverage by 11.66\%, and runs 2.76× faster, highlighting its scalability and effectiveness for directed fuzzing in complex hardware systems.\\
\end{abstract}


\section{Introduction}
\vspace{1em}
Modern hardware systems, from resource-constrained compact edge devices to complex, heterogeneous platforms, are rapidly evolving to support the increasing demands of next-generation applications. These designs often integrate billions of transistors and a wide range of components, making the hardware verification process both critical and increasingly challenging. Despite significant advances, traditional verification techniques, including functional and formal methods \cite{formal_ver_survey,formal_ver_survey_2,clip}, struggle to scale with the growing complexity of hardware designs to expose subtle or potential vulnerabilities. Recent advancements in leveraging Large Language Models in automating bug identification and mitigation \cite{paria2023divas,pearce,lasa,veridebug,dispel,marvel,spell} have shown promise, but these techniques remain limited in their scope and accuracy, paving the way for fuzzing as a complementary approach to uncover deeper, hard-to-find bugs. Fuzzing is an automated testing method that repeatedly generates inputs, often randomly or through guided mutation, to expose bugs, crashes, or unexpected behaviors~\cite{microsoftrisk}.  Inspired by well-established software fuzzing techniques, hardware fuzzing techniques emerged as a promising verification approach~\cite{Laeufer'18, Li'21, Hur'21, Trippel'22, Muduli'20, Kabylkas'21, Kande'22,saravanan2025synfuzzleveragingfuzzingnetlist,Saravanan'24a}. Coverage-Guided Graybox Fuzzing (CGF)~\cite{Laeufer'18} has become a foundational technique in this space, and RTL-level Fuzzing frameworks like~\cite{Kande'22, hybrid, Hur'21, Canakci'23} have demonstrated effectiveness in achieving broad coverage, particularly across designs with complex datapaths~\cite{RISC-V, Morlkx}. In CGF, input generation is directed by feedback from real-time code coverage, allowing the fuzzer to prioritize input mutations that reveal unexplored paths in the design to improve functional coverage of RTL designs. While CGF-based techniques mark a significant advancement, they fall short in addressing the practical realities of hardware design workflows. Unlike software, hardware is typically developed incrementally, with new modules added or existing ones modified across design iterations. In such flows, verification efforts must focus on specific changes rather than re-testing the entire system. Directed Graybox Fuzzing (DGF) has been proposed as an alternative, which, unlike CGF, enables targeted fuzzing by steering test generation toward specific regions of interest. This is particularly useful in scenarios such as patch validation, bug localization, or module-specific testing. DirectFuzz~\cite{direct}, a DGF-based implementation, aimed to bring directionality into the fuzzing process to improve targeted coverage.
Despite its innovation, DirectFuzz exhibits several key limitations, including (i) inability to capture hardware-specific semantics and structure, (ii) misaligned coverage metrics that do not align well with established verification practices, (iii) limited scalability when targeting multiple design regions, and (iv) 
an inability to perform targeted verification across multiple modules. These challenges motivate the need for a more scalable and structurally aware directed fuzzing framework tailored to the realities of modern hardware design and verification.

In this work, we present \profuzz, a novel hardware fuzzing framework based on DGF, specifically designed to overcome the challenges, such as abstraction mismatches, limited coverage precision, and poor scalability. Unlike existing methods, \profuzz~operates directly at the hardware’s native abstraction level, allowing it to accurately model inherent hardware behaviors and utilize coverage metrics that reflect hardware-specific characteristics. Additionally, our framework seamlessly integrates with industry-standard Electronic Design Automation (EDA) tools, facilitating easy adoption within established design and verification workflows. 

Our main contributions are summarized as follows:

\begin{itemize}
\item \textbf{Hardware-native Directed Graybox Fuzzing framework:} Captures intrinsic hardware characteristics and delivers efficient coverage scaling with increasing complexity of target regions.
\item \textbf{Scalable target site selection strategy:} Supports both random and cost-driven approaches, adaptable to diverse hardware designs.
\item \textbf{Integration of Automatic Test Pattern Generation (ATPG):} Enables directed seed generation for target sites, utilizing conflict-aware merging to enhance seed quality and mutation diversity.
\item \textbf{Cross-module verification support:} Employs automated submodule generation to enable comprehensive fuzzing across multiple design regions.
\end{itemize}

\section{Background}\label{background}
\vspace{0.5em}

\subsection{Fuzzing in Hardware: CGF and DGF}\label{fuzzing}

\begin{figure}[!htbp]
\vspace{-1em}
  \centering
  \includegraphics[width=\columnwidth]{
  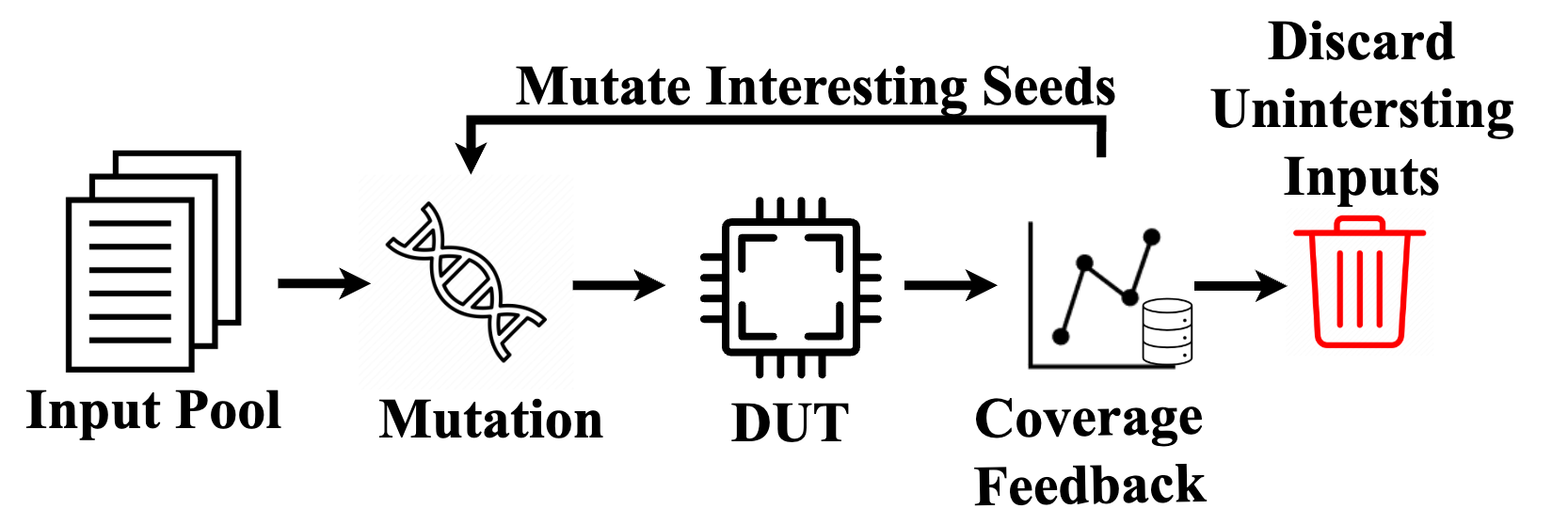} \vspace{-2em}
  \caption{Overview of Coverage Graybox Fuzzing (CGF).
  } \label{fig:cgf}
\end{figure}

Hardware fuzzing \cite{Kande'22, Trippel'22, hybrid} is a dynamic testing technique that exposes hidden bugs and vulnerabilities by subjecting the Design Under Test (DUT) to randomized input stimuli. Instead of relying purely on random inputs, most hardware fuzzers employ CGF, which uses coverage feedback to steer input mutations and efficiently explore the design’s state space, as illustrated in Fig. \ref{fig:cgf}. The process starts with generating an initial set of input seeds, which are then iteratively mutated based on real-time coverage data collected from the DUT. Popular mutation strategies, borrowed from the American Fuzzy Lop (AFL) framework~\cite{AFL'23}, include bit-flipping, byte swapping, and arithmetic operations. These techniques systematically modify inputs to exercise diverse execution paths and trigger rare hardware states. RFUZZ \cite{Laeufer'18} is among the pioneering hardware fuzzing frameworks that apply CGF to maximize code coverage in hardware designs. By using coverage feedback, RFUZZ iteratively mutates inputs to accelerate coverage growth. However, a broad exploration strategy of CGF-based techniques often results in inefficient use of computational resources. More importantly, it lacks the targeted focus necessary to validate specific modules or recent incremental changes, limiting its applicability for hardware development workflows that rely heavily on iterative design updates. 

\begin{figure}[htb!]
  \centering
  \includegraphics[width=0.9\columnwidth]{
  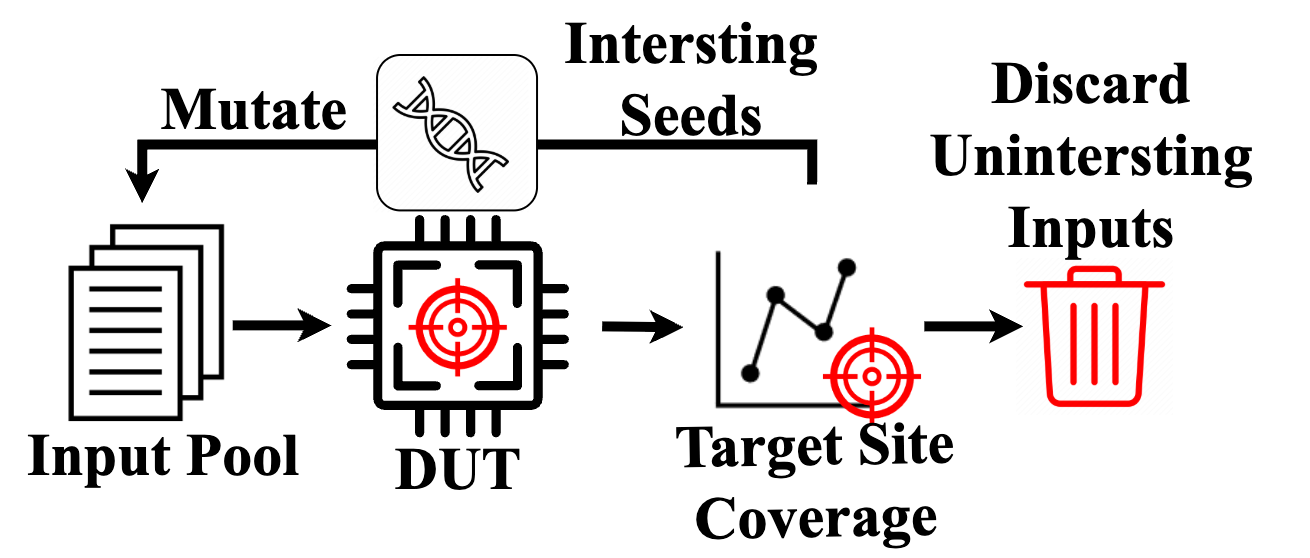} \vspace{-1em}
  \caption{Overview of Directed Graybox Fuzzing (DGF).
  } \label{fig:dgf}
\end{figure}

In contrast, DGF focuses on specific target sites within hardware designs by monitoring coverage at these sites during fuzzing and applying mutations to input seeds based on improvements in targeted coverage, as illustrated in Fig. \ref{fig:dgf}. The first practical implementation of DGF for hardware, DirectFuzz \cite{direct}, aims to maximize coverage in designated regions of a design. It achieves this by translating RTL code into a software model using Verilator \cite{Verilator}, treating multiplexer selection signals similarly to software branches to serve as coverage points. DirectFuzz further builds a module connectivity graph to steer input generation toward activating these signals. While this approach effectively targets specific hardware sites, it has notable drawbacks, especially its dependence on software-level abstractions and limited scalability.

\begin{table*}[!ht]
\centering
\caption{Comparative analysis of the proposed \profuzz~approach with existing CGF and DGF solutions}
\label{tab:background}
\resizebox{\textwidth}{!}{%
\begin{tabular}{c|c|c|c|c|c|c|c}
\hline
\textbf{Technique}  & \textbf{Abstraction} & \textbf{Method} & \multicolumn{1}{c|}{\textbf{Input}} & \multicolumn{1}{c|}{\textbf{Simulator}} & \multicolumn{1}{c|}{\begin{tabular}[c]{@{}c@{}}\textbf{Iterative} \\ \textbf{Approach?}\end{tabular}} & \multicolumn{1}{c|}{\begin{tabular}[c]{@{}c@{}}{\textbf{\#Target Sites}} \\ \textbf{(Min - Max)}\end{tabular}}& \multicolumn{1}{c}{\begin{tabular}[c]{@{}c@{}}\textbf{EDA Toolflow} \\ \textbf{Integration?}\end{tabular}} \\ \hline
RFUZZ \cite{Laeufer'18}     
& FIRRTL      
& CGF    
& Series of bits                          
& Verilator                              
& \xmark                                 
& N/A\btrs                                                                              
& \xmark                                                                               
\\ \hline
DifuzzRTL \cite{Hur'21}    
& RTL (SW)    
& CGF    
& Assembly                            
& Verilator                               
& \xmark                                 
& N/A\btrs                                                                                
& \xmark                                                                      
\\ \hline
TheHuzz  \cite{Kande'22}    
& RTL (HW)    
& CGF    
& Assembly                            
& any\blzs                              
& \xmark                                 
& N/A\btrs                                                                                
& \cmark
\\ \hline
ProcessorFuzz \cite{Canakci'23}  
& RTL (SW)    
& CGF    
& Assembly                           
& Verilator                               
& \xmark                                 
& N/A\btrs                                                                               
& \xmark
\\ \hline
DirectFuzz  \cite{direct}
& RTL (SW)    
& DGF    
& Series of bits                            
& Verilator                               
& \xmark                                 
& 6 - 107                                                                               
& \xmark                                                                                
\\ \hline
\profuzz~(this work)
& RTL/GLN  (HW)        
& DGF    
& Series of bits                            
& any\blzs                            
& \cmark                                 
& 81 - 2857                                                                              
& \cmark                                                                                
\\ \hline
\end{tabular}%
}
\footnotesize
\raggedright \\
{\vspace{0.75mm}
\btrs \textbf{N/A}: Not applicable since CGF solutions do not rely on targeted sites but rather follow broad exploration. \\
\blzs \textbf{any}: Any commercial industry-standard HDL simulator.}

\vspace{-1em}
\end{table*}

\subsection{Categorizing Hardware Fuzzing Frameworks}\label{category}

Existing hardware fuzzing frameworks can be broadly classified into three categories:

\paragraph{\textbf{Direct Use of Software Fuzzers on Hardware}}
This method applies software fuzzers like AFL~\cite{AFL'23} directly to hardware designs, generating randomized inputs and using software-oriented coverage metrics (e.g., branch coverage) to guide the fuzzing process.

\paragraph{\textbf{Fuzzing Hardware as Software}}
Hardware designs written in RTL languages such as Verilog or VHDL are translated into software models using tools like Verilator. These software abstractions are then fuzzed with standard software fuzzers.

\paragraph{\textbf{Fuzzing Hardware as Hardware}}
This approach performs fuzzing directly on hardware representations, such as RTL or gate-level netlists, leveraging industry-standard EDA tools like Synopsys VCS \cite{Synopsys-VCS} or Cadence Xcelium \cite{Cadence-Xcelium}. These tools provide native hardware coverage metrics, including branch, toggle, FSM, expression, and line coverage, aligned with conventional hardware verification flows.

\subsection{Prevailing Challenges of Directed Gray-box Fuzzing} \label{challenges}

\subsubsection{Software-Level Abstraction}

Translating RTL to C++ through Verilator creates a mismatch by failing to capture key hardware semantics like concurrent execution, precise timing constraints, register behavior, and signal interactions. This abstraction gap means the software model doesn’t fully represent the actual hardware behavior, limiting the effectiveness of fuzzing based on it.

\begin{tcolorbox}
  \textbf{Challenge C1:} The translated software version of the hardware by Verilator fails to account for the inherent hardware characteristics.
\end{tcolorbox}

\subsubsection{Misaligned Coverage Metrics}

DGF relies on coverage feedback derived from the software model generated by Verilator rather than from the hardware itself. While some software constructs like branches loosely correspond to hardware elements such as multiplexers, this mapping is limited. Important hardware coverage aspects, like signal toggling, FSM transitions, expression evaluation, and block activation, are either lost or poorly represented. This abstraction gap leads to coverage metrics that do not accurately reflect true hardware behavior, causing the fuzzing process to focus on inputs that improve software-like coverage while missing other critical hardware states.

\begin{tcolorbox}
  \textbf{Challenge C2:} The coverage metric fails to capture intrinsic hardware characteristics associated with the actual hardware design.
\end{tcolorbox}

\subsubsection{Lack of Scalability}

Current DF approaches are designed to handle $<$100 target signals limited to specific structures like multiplexers (MUX), which limits the scalability as target regions grow. It also lacks flexibility to handle important control or datapath signals beyond MUX logic, restricting its ability to support broader and more complex verification tasks.

\begin{tcolorbox}
\textbf{Challenge C3:} Current DGF techniques like DirectFuzz target fewer than 100 signals, limiting their scalability and failing to cover broader target regions.
\end{tcolorbox} \label{C3}

\subsubsection{Lack of Cross-Module Verification}

Current DFG approaches like DirectFuzz \cite{direct} focus on individual modules, such as targeting only the RX or TX block in a UART design. While effective for isolated testing, this approach falls short when verifying interactions across multiple modules. Since many real-world bugs stem from inter-module dependencies, the inability to validate cross-module behavior is a critical limitation for comprehensive hardware verification.

\begin{tcolorbox}
\textbf{Challenge C4:} Current DGF techniques like DirectFuzz restrict the fuzzing to single target module at a time, making it ineffective for verifying interactions across multiple interconnected components.
\end{tcolorbox}

Table I summarizes existing hardware fuzzing approaches, categorizing them by abstraction level and core methodology, while highlighting the need for our proposed approach to address the limitations of current techniques.\\

\begin{figure}[h]
\vspace{-1em}
  \centering
  \includegraphics[width=1 \linewidth]{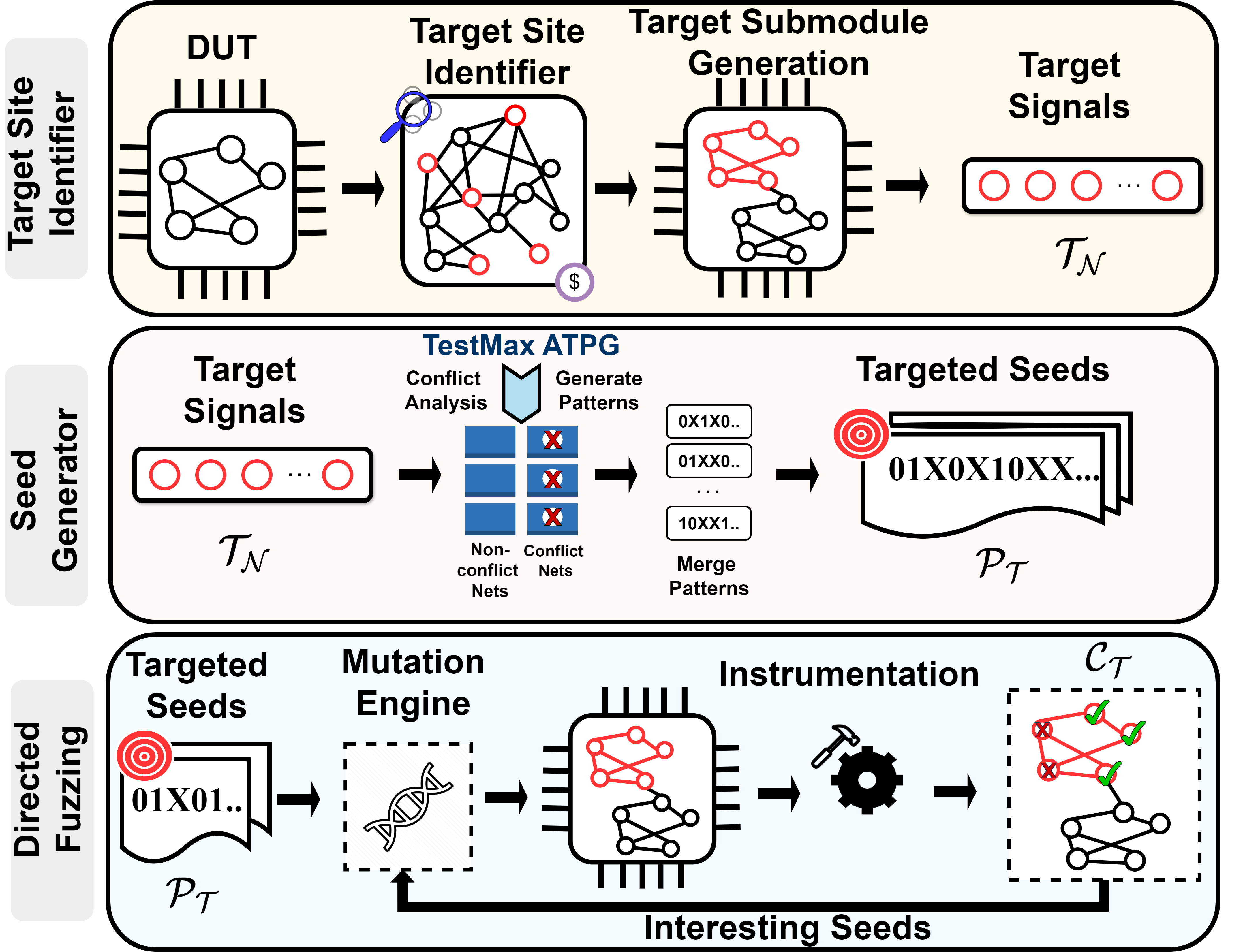}
  \caption{Overview of \profuzz~methodology highlighting major steps.} 
  \label{fig:profuzz}
 \vspace{-1em}
\end{figure}

\section{PROFuzz Methodology} 
\label{profuzz}

Fig. \ref{fig:profuzz} demonstrates the overview of the \profuzz~methodology, highlighting three major components: (1) Target Site Identifier, (2) Seed Generator, and (3) Directed Fuzzing. The input to the Target Site Identifier is the hardware design ($\mathcal{I_N}$) at its native abstraction level (RTL or gate-level). \profuzz~constructs a hypergraph $G = (\mathcal{V}, \mathcal{E})$ from the netlist abstraction (for RTL, it performs initial synthesis to generate the gate-level netlist), where $\mathcal{V}$ represents the set of gates and $\mathcal{E}$ the interconnections between them. Using configuration parameters ($\theta$), the Target Site Identifier selects a subset of nodes ($\mathcal{T_N}$) for verification and coverage improvement. Next, \profuzz~applies conflict-PI (Primary Input) and fan-in cone analysis to identify non-conflicting nets and leverages ATPG to generate activation patterns. These patterns are merged
to create a targeted seed pattern ($\mathcal{P_T}$) that maximizes mutation potential. The directed fuzzing loop then iteratively mutates inputs, monitors coverage at target sites ($\mathcal{C_T}$), and updates the seed pool until a coverage goal or timeout is reached. The following sections provide a detailed overview of each stage.

\subsection{Target Site Identifier} \label{ts}

\profuzz~generates seed inputs to maximize coverage of target sites in hardware designs (Fig. \ref{fig:profuzz}). Unlike DirectFuzz, which uses signals from C++-translated models, \profuzz~operates directly on native hardware abstractions, preserving design semantics and addressing challenges \textbf{C1} and \textbf{C2}.
Target signals ($\mathcal{T_N}$) can be selected manually or automatically. In the manual approach, the verification engineer selects signals based on specific verification requirements. For automatic selection, a cost function based on structural (fan-in $\mathbb{FI}$, fan-out $\mathbb{FO}$) and stochastic properties (Shannon entropy $\mathbb{H}$) is used to select target signals. This function is applied to each edge $i \in \mathcal{E}$ as follows:

\begin{equation}
\mathcal{C}_i = {cost\_fun(\mathbb{FI}_i, \mathbb{FO}_i, \mathbb{H}_i)}, \forall i \in \mathcal{E}
\label{eq1}
\end{equation}
The nets/signals are ranked and selected based on $\mathcal{C}_i$ values along with heuristics such as controllability, observability, and topological depth to maximize verification efficiency.

\setlength{\textfloatsep}{0pt} 
\begin{algorithm}[!htb]
\caption{\profuzz~Methodology}
\label{algo:profuzz}
\DontPrintSemicolon 
  \SetKwInOut{Input}{Input}
  \SetKwInOut{Output}{Output}
  \Input{Gate-level Netlist ($\mathcal{I_N}$), config ($\theta$)}
  \Output{Simulation Vectors ($\mathcal{S_I}$), Target Site Coverage ($\mathcal{C_T
  }$)}

  $G$ $\leftarrow$ \text{parse $\mathcal{I_N}$ and construct hypergraph} $: G = \{\mathcal{V},\mathcal{E}\}$\\
  $G^T$ $\leftarrow$ \text{\textit{topological\_sort}($G$)}\\
  \tcc{\noindent\hspace{\algorithmicindent} {\hspace{0.5cm} \texttt{Target Site Identifier} \hspace{0.5cm}}}
  $\mathcal{T_N}$ $\leftarrow$ \text{\textbf{target\_net\_selection}($G,\theta$)}\\
  \tcc{\noindent\hspace{\algorithmicindent} {\hspace{1cm} \texttt{Seed Generator} \hspace{0.5cm}}}
  $\mathcal{P_T}, \mathcal{N}_0, \mathcal{N}_1$ $\leftarrow$ \text{\textbf{pattern\_generation}($G,G^T,\mathcal{T_N},\theta$)} \hspace{2cm}
  //\hspace{2cm}$\triangle$ \textit{Detailed in Algo 2}\\
  $\mathcal{S_T}$    $\leftarrow$  $\mathcal{P_T}$,
  $\mathcal{C_T} \leftarrow 0$, $\mathcal{C_T^{\text{init}}} \leftarrow 0$ \\
  \tcc{\noindent\hspace{\algorithmicindent} {\hspace{1cm} \texttt{Directed Fuzzing} \hspace{0.5cm}}}

  \While{termination criteria not met}{
    \ForEach{seed $S \in \mathcal{S_T}$}{
        
        $\mathcal{S_M} \leftarrow$ \textit{$mutate$}($S$)\;
        $\mathcal{C_T} \leftarrow$ \textit{$run$}($\mathcal{I_N}$, $\mathcal{S_M}$) \\
        
        \If{$\mathcal{C_T} > \mathcal{C_T^{\text{init}}}$}{
            $\mathcal{S_I} \leftarrow \mathcal{S_I} \cup \{S_M\}$ \\ 
            $\mathcal{S_D} \leftarrow \mathcal{S_M} $\\
            $\mathcal{C_T^{\text{init}}} \leftarrow \mathcal{C_T}$ \\
        }
    }
    \textbf{until} \text{time or coverage threshold}
} 
\Return{$\mathcal{S_I}, \mathcal{C_T}$}
 
\end{algorithm}

Existing approaches, as noted in Section \ref{challenges}, struggle with scalability when handling many target signals, while selecting too many can add significant overhead, resulting in degraded performance. To address this, \profuzz~creates a lightweight target submodule that includes only the selected signals. This submodule is integrated into the original design, enabling precise monitoring without full DUT instrumentation, thus solving Challenge \textbf{C3}.
This method reduces overhead and supports cross-module verification, addressing Challenge \textbf{C4}. The parameter ($\theta$) in Algorithms \ref{algo:profuzz} and \ref{algo:atpg} configures the selection. Signals within the target submodule are labeled `target' and the ATPG engine focuses on maximizing coverage for these nodes ($\mathcal{T_N}$).

\setlength{\textfloatsep}{0pt} 
\begin{algorithm}[!htb]
\caption{Pattern Generation}
\label{algo:atpg}
\DontPrintSemicolon 
  \SetKwInOut{Input}{Input}
  \SetKwInOut{Output}{Output}
  \Input{$G,G^T,\mathcal{T_N},\theta$}
  \Output{$\mathcal{P_T}, \mathcal{N}_0, \mathcal{N}_1$}
  $ID_{N}$ $\leftarrow$ \text{\textbf{get\_id}($G,G^T,\mathcal{T_N}$)}\\
  $\mathcal{P}_{all}$ $\leftarrow$ \text{\textbf{gen\_non\_conflict\_patt}($G,\mathcal{T_N}, ID_{N}$)}\\
  $\mathcal{P}_{merged}, \mathcal{N}_0, \mathcal{N}_1, \mathcal{N}_C$ $\leftarrow$ \text{\textbf{merge\_pattern}($\mathcal{P}_{all}$)}\\
  $\mathcal{T_N} \leftarrow \mathcal{T_N} - \mathcal{N}_C$ \\
  \If{$\theta.submodule$ is $True$}
  {
       $ID_{N}^M$ $\leftarrow$ \text{\textbf{get\_id}($G,G^T,\mathcal{T_N^M}$)}\\
       $\mathcal{P}_{all}^M$ $\leftarrow$ \text{\textbf{gen\_non\_conflict\_patt}($G,\mathcal{T_N^M}, ID_{N}^M$)}\\
       $\mathcal{P}_{merged}^M, \mathcal{N}_0, \mathcal{N}_1, \mathcal{N}_C$ $\leftarrow$ \text{\textbf{merge\_pattern}($\mathcal{P}_{all}^M$)}\\
       $\mathcal{T_N} \leftarrow \mathcal{T_N^M} - \mathcal{N}_C$ \\
       $\mathcal{P_T} \leftarrow \mathcal{P}_{merged}^M$\\
       \textbf{return} $\mathcal{P_T}, \mathcal{N}_0, \mathcal{N}_1$
  }
  $\mathcal{P_T} \leftarrow \mathcal{P}_{merged}$\\
  \textbf{return} $\mathcal{P_T}, \mathcal{N}_0, \mathcal{N}_1$
 
\end{algorithm}

\subsection{Seed Generator} \label{atpg}

After identifying the target nets ($\mathcal{T_N}$), \profuzz~generates seed patterns to maximize coverage of target sites during fuzzing. It integrates Synopsys TestMAX ATPG to create hierarchical IDs and activation patterns (0 and 1) for each net $n_i \in \mathcal{T_N}$ using the stuck-at-fault model.
\profuzz~performs conflict-PI and fan-in-cone analysis to identify conflicting nets ($\mathcal{N}C$) and merges non-conflicting patterns into a single seed pattern ($\mathcal{P}{merged}$). The merging maximizes \textit{Don't Care (X)} values for Primary Inputs (PIs), increasing the number of potential mutations. Based on the merged patterns, nets are assigned to $\mathcal{N}_0$ (for 0) or $\mathcal{N}_1$ (for 1). For target submodules, \profuzz~iteratively generates seed patterns for both the submodule and the main module. Algorithm \ref{algo:atpg} outlines this ATPG-based pattern generation.

\subsection{Directed Fuzzing} \label{fuzz}

The input design ($\mathcal{I_N}$) is fuzzed using targeted seed patterns ($\mathcal{P_T}$), forming the initial seed pool ($\mathcal{S_T}$) to stimulate the target submodule. As shown in Fig. \ref{fig:profuzz}, the coverage points correspond to the target nodes ($\mathcal{T_N}$), which represent the signals selected for focused verification. To track fuzzing effectiveness, \profuzz~uses industry-standard EDA tools to analyze the design and measure coverage ($\mathcal{C_T}$) at these target nodes. The fuzzing process follows an iterative feedback loop: mutations are applied to each seed to generate new test vectors ($\mathcal{S_M}$), and only mutations that increase coverage are retained to update the seed pool. We use the AFL-style mutation algorithm \cite{Laeufer'18}. The fuzzing campaign ends when a predefined coverage threshold is met or the allocated time expires, ensuring efficiency and controlled resource usage.

\begin{figure*}[!htbp]
  \centering
  \includegraphics[width=0.8\textwidth]{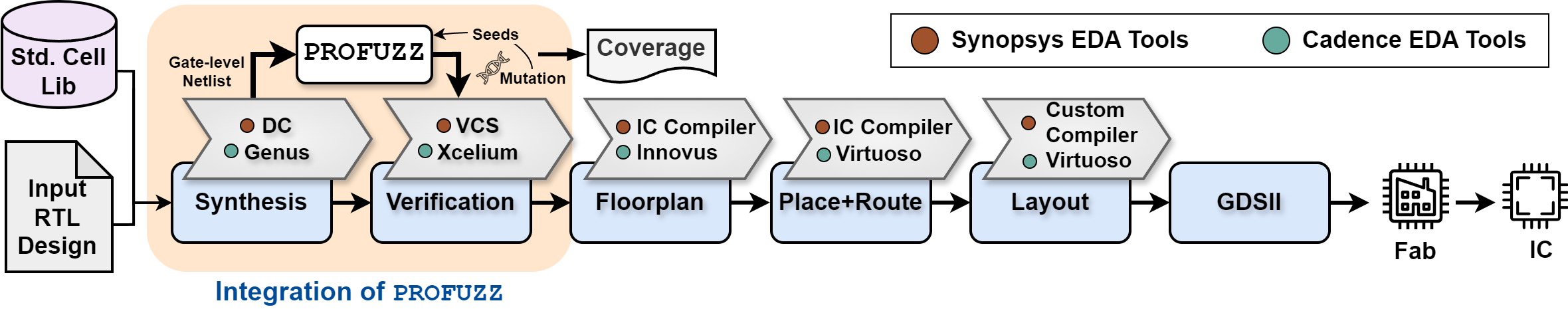}
  \vspace{-1em}
  \caption{Integrating \profuzz~into the commercial EDA tool flow.} 
  \label{fig:eda_profuzz}
\end{figure*}

\subsection{Integration into commercial EDA tool flow} \label{eda_toolflow}

\profuzz~is fully compatible with the standard commercial EDA tool flow for ASICs and can be seamlessly integrated into existing verification and simulation environments for performing directed fuzzing for any hardware design. This compatibility ensures that \profuzz~can be readily applied to a wide range of hardware designs to enhance functional coverage and uncover potential implementation bugs. Fig. \ref{fig:eda_profuzz} demonstrates the integration of \profuzz~into the commercial (Synopsys/Cadence) EDA tool flow.

\section{Experimental Evaluation}

\subsection{Experimental Setup}
\profuzz~is fully compatible with the industry-standard IC design and verification flow, enabling seamless Directed Fuzzing. For hardware design simulations, we use Cadence Xcelium, a widely adopted EDA tool in both industry and academia. Target state and site coverage metrics are extracted using Cadence Integrated Metrics Center (IMC) through custom TCL and Python scripts. Seed patterns for activating target sites are generated using Synopsys TestMAX (V-2023.12-SP5).
Experiments were conducted on an Intel Core i7 (2.4 GHz) running a Linux OS. The fuzzing campaign continued until coverage reached $\ge$ 90\% (configurable threshold).

\subsection{Evaluation Dataset}
We evaluate \profuzz~across a diverse set of hardware designs as illustrated in Table \ref{tab:benchmarks}. For each of the designs, we assess the effectiveness of our proposed framework by performing target site selection under two configurations: (1) random selection and (2) Cost Function-based selection. For the cost function-based configuration, we selected target nodes using two criteria: (i) nodes with a cost function value $\mathcal{C}_i$ greater than 0.3 and (ii) nodes within the top 10 percentile of the ranked distribution.

We evaluate \profuzz~across a range of hardware designs, as shown in Table \ref{tab:benchmarks}. For each design, we assess its effectiveness by performing target site selection in two configurations: (1) random selection and (2) Cost Function-based selection. In the cost function-based configuration, target nodes are selected based on two criteria: (i) nodes with a cost function value $\mathcal{C}_i$ $>$ 0.3, and (ii) nodes within the top 10 percentile of the ranked distribution.

\begin{table}[!ht]
\centering
\caption{Details of evaluation benchmarks}
\label{tab:benchmarks}
\resizebox{\columnwidth}{!}{%
\begin{tabular}{c|ccc|c}
\hline
\multicolumn{1}{c|}{\textbf{Design}} & \textbf{Input Width} & \textbf{\#Signals} & \textbf{\#Gates\blzs} & \textbf{Description} \\ 
\hline
ORCTRL                       &  108    &  3328    &     3383    &    Control Unit of OR1200 CPU       \\ \hline
GPIO                         &   70   &    905  &     834    &  General Purpose Input/Output           \\ \hline
FIR                          &    33  &    3222  &     2521    &    Finite Impulse Response Filter (DSP)         \\ \hline
IIR                          &  33    &  2801    &     2545    &    Infinite Impulse Response Filter (DSP)       \\ \hline
PWM                          &   54   &   3454   &    3402     &   Pulse Width Modulation          \\ \hline
SPI                          &    102  &   7617   &  7447       &   Serial Peripheral Interface           \\ \hline
I2C                          &   18   &    2125  &      2121   &   Inter-Integrated Circuit          \\ \hline
UART                         &   28   &   1522   &     1369    & Universal Async.  Receiver/Transmitter \\ \hline
PicoRV32                         &  101    &   53306   &     53215    &   RISC-V Processor          \\ \hline
DFT                          &    63  & 189567     &      187168    &   Discrete Fourier Transform (DSP)          \\ \hline
IDFT                         &    63  &    190361  &     187986     &   Inverse DFT (DSP)          \\ \hline
DES3                         &   235   &    3807  &      3630   &  Triple DES Encryption           \\ \hline
SHA256                       & 515     & 8045      &       7884  &   Secure Hash Algorithm 256-bit         \\ \hline
MIPS                       &   73   &  26694    &    25532     &   $\mu$P w/o Interlocked Pipeline Stages         \\ \hline
Sodor1                      &   72   &   13692   &     13394    &   RISC-V simplified $\mu$P with $X$ stages         \\ \hline
Sodor3                     &   72   &   13991   &     13652    &   RISC-V simplified $\mu$P with $X$ stages         \\ \hline
\end{tabular}%
}
\scriptsize
\raggedright \\
\blzs \textbf{\#Gates} include the count of combinational and sequential logic gates. All the designs are synthesized using gscl45nm standard cell library. 
\vspace{1em}
\end{table}

\subsection{Evaluation Results}

Table \ref{tab:v1_full} presents the coverage (in \%) and the runtime (in seconds) for the evaluation benchmarks with randomly selected target signals.
\profuzz~demonstrates high efficiency on datapath-centric designs such as \texttt{FIR} and \texttt{IIR}, achieving 100\% coverage while targeting only 32.5\% and 15.1\% of internal signals, respectively, in under 5 seconds. \texttt{GPIO} attains 99.39\% coverage in 2.36 seconds despite a moderate target ratio (35.9\%). For more complex control-heavy designs like \texttt{ORCTRL}, \profuzz~achieves 94.54\% coverage while targeting just 12.5\% of internal nodes, demonstrating strong robustness. In contrast, designs such as \texttt{PWM} and \texttt{SPI} require over 50\% of internal nodes to be targeted, achieving 53.24\% and 75.63\% coverage, respectively, reflecting the increased difficulty of covering dense target sets; nonetheless, \texttt{SPI} reaches 75.63\% coverage in only 5.18 seconds. The \texttt{UART} design targets 64.6\% of internal signals and achieves 93.81\% coverage in 15.21 seconds, illustrating scalability to medium-complexity modules. Cryptographic cores (\texttt{DES3}, \texttt{SHA256}) achieve over 98\% coverage within short runtimes, indicating strong observability despite algorithmic complexity. For \texttt{I2C}, with 1209 target nodes, the proposed framework achieves 94.10\% coverage in approximately 240 seconds, further highlighting scalability of \profuzz. However, for large designs such as \texttt{PicoRV32}, DFT, and IDFT, full instrumentation incurs significant overhead (\btrs) and resource usage (in Table~\ref{tab:v1_full}), limiting practicality. To address this, we employ a target submodule generation strategy, with results in Table~\ref{tab:v2_sub}, enabling scalable analysis of large-scale designs.

\begin{table}[!ht]
\centering
\caption{Evaluation for Random Selection of Target Signals }
\label{tab:v1_full}
\vspace{-0.5em}
\resizebox{\columnwidth}{!}{%
\begin{tabular}{c|c|c|c}
\hline
\textbf{Design} & \textbf{\#Target Signals\nlbs} & \textbf{Coverage}       & \textbf{Time (s)}          \\ \hline
ORCTRL &           415            & 94.54\%           & 20.41            \\ \hline
GPIO   &           325            & 99.39\%           & 2.36             \\ \hline
FIR    &      1049           & 100\%          &        4.81             \\ \hline
PWM    &           2008           & 53.24\%           & 10.38            \\ \hline
SPI    &           2857           & 75.63\%           & 5.18             \\ \hline
I2C    &           1209           & 94.10\%           & 240.37             \\ \hline
UART   &           983            & 93.81\%           & 15.21            \\ \hline

PicoRV32   &           5287           & Ins Overhead\btrs & Ins Overhead\btrs \\ \hline

IIR    &       423            & 100\%          & 4.31             \\ \hline
 
DFT    &   9145     & Ins Overhead\btrs   & Ins Overhead\btrs   \\ \hline

IDFT   &   9518    & Ins Overhead\btrs   & Ins Overhead\btrs   \\ \hline

DES3   &       470            & 98.62          & 3.64             \\ \hline
SHA256 &      993            & 98.09          & 10.09         \\ \hline
\end{tabular}
}
\scriptsize
\raggedright \\
\btrs \textit{Ins Overhead} refers to instrumentation overhead, implying additional resource usage incurred due to monitoring target signals during the fuzzing process.\\
\nlbs \textit{\#Target Signals} denotes the number of signals selected randomly from the design, ranging between 2\%-50\% of the total signals.
\end{table}

Table \ref{tab:v2_sub} presents the coverage values (in \%) for different benchmarks with target submodule configuration and respective time (in seconds), demonstrating the ability of \profuzz~to isolate and verify signal groups that span multiple modules and instances. The instrumentation overhead for large designs, such as \texttt{PicoRV32}, is effectively mitigated using the target submodule generation strategy, which enables directed seed generation focused on specific regions, as presented in Table \ref{tab:v2_sub}. For \texttt{PicoRV32}, with 455 target nodes, \profuzz~achieves 60.42\% coverage, while for \texttt{ORCTRL} it achieves similar coverage (with 415 nodes) under 20 seconds, but a more deeply embedded subset of 200 nodes requires nearly 300 seconds, highlighting the increased effort to activate nested logic. Smaller modules like \texttt{GPIO}, \texttt{FIR}, and \texttt{I2C} show consistent performance due to easier access to target sites. \texttt{UART} achieves 95.63\% coverage for 219 target nodes, comparable to the full-module result of 93.81\% with 983 targets (Table~\ref{tab:v1_full}), further validating the submodule approach.

\begin{table}[!htbp]
\centering
\caption{Evaluation for Generated Target Submodule}
\label{tab:v2_sub}
\resizebox{\columnwidth}{!}{%
\begin{tabular}{c|c|c|c|c} 
\hline
\textbf{Design} & \textbf{\begin{tabular}[c]{@{}c@{}}\#Sub\_Target \\Signals\blzs\end{tabular}} & \textbf{\begin{tabular}[c]{@{}c@{}}\#Main\_Target \\Signals\btrs\end{tabular}} & \textbf{Coverage} & \textbf{Time (s)} \\ \hline
ORCTRL &           200      &  749    & 92.23\%     & 300.51  \\  \hline
GPIO   &           416      &  209    & 99.94\%     & 2.47    \\  \hline
FIR    &           587      &    882   & 100\%    & 2.63    \\ \hline
PWM    &           710      &  975    & 55.38\%     &    11.43   \\ \hline
I2C    &           437      &    597  & 93.15\%     & 240.37 \\ \hline
UART   &           219      &   189       & 95.63\%     & 2.59   \\  \hline
DFT    &           307      &  12454    & 97.84\%     & 60.71 \\ \hline
IDFT    &          387       &  12749 & 96.86\%     &  75.12    \\ \hline
PicoRV32   &           455      &   5189   & 60.42\%       & 636.24 \\  \hline
MIPS   &          268       &   3129  & 52.48\%   & 620.12  \\ \hline
Sodor1Stage   &      707           &  2891   & 59.36\%   & 660.57   \\ \hline
Sodor3Stage   &        680         &  3003   & 61.27\%  & 710.68   \\ \hline
\end{tabular}
}
\scriptsize
\raggedright \\
\blzs\#Sub\_Target Signals: target signals encapsulated inside the target submodule.\\
\btrs\#Main\_Target Signals: signals in the original design connecting the target submodule.
\end{table}

\begin{figure}[!htbp]
  \centering
  \includegraphics[width=\columnwidth]{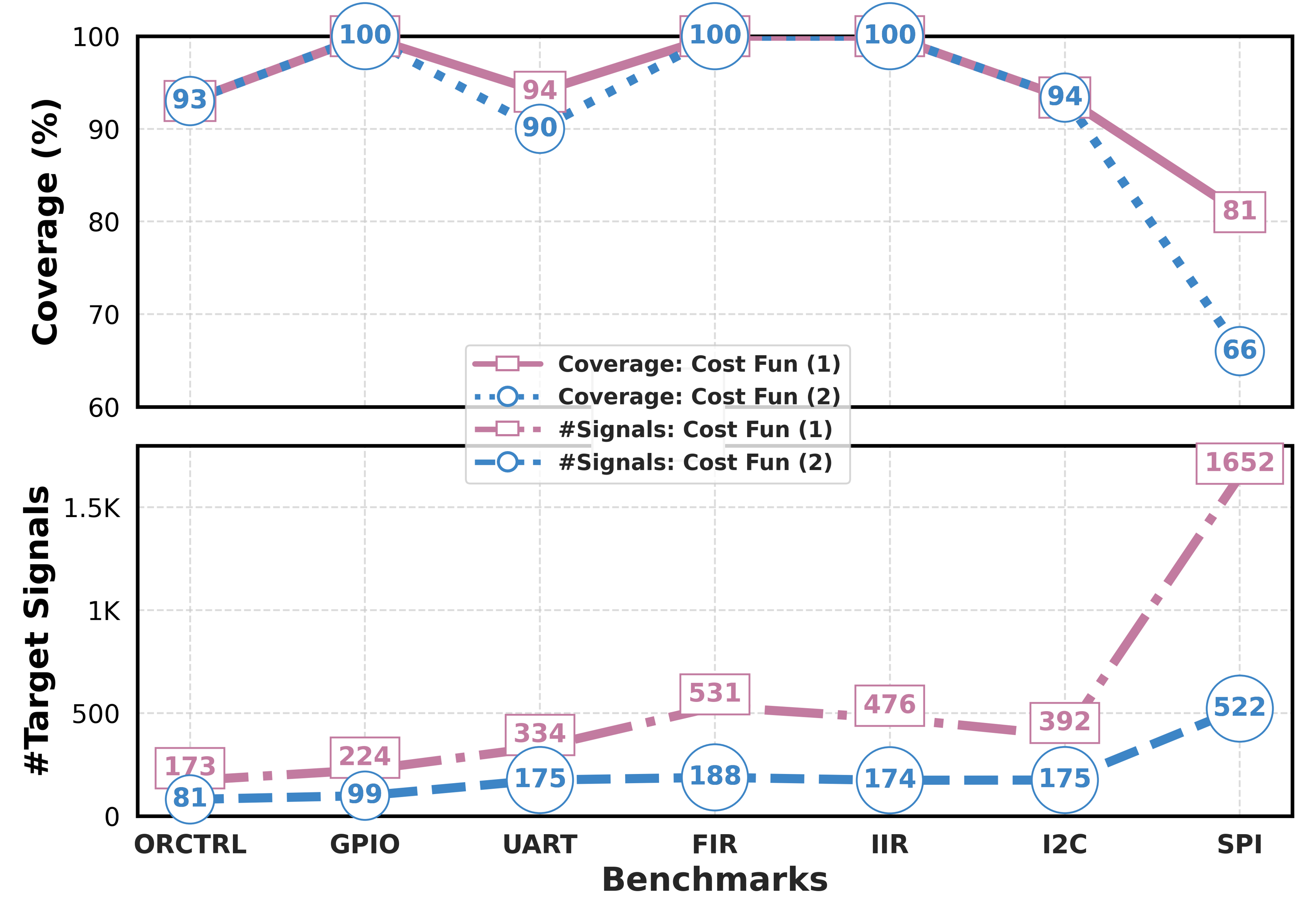} 
  \caption{Coverage(\%) and the number of target signals (\#Target Signals) for seven benchmarks with varying sizes where the target sites are selected based on cost functions: (1) $\mathcal{C}_i \ge 0.3$ and (2) $\mathcal{C}_i \in P_{10}$.} 
  \label{fig:cost_fun}
\end{figure}

Fig. \ref{fig:cost_fun} shows the coverage (\%) and number of target signals for various benchmarks selected using two different cost functions: (1) nets with $\mathcal{C}_i \geq 0.3$ and (2) nets in the top 10 percentile ($P_{10}$) of the ranked cost function distribution. \profuzz~achieves over 90\% coverage across all benchmarks except \texttt{SPI}. The number of target signals is also highlighted to demonstrate \profuzz's scalability. Notably, coverage remains consistent across both cost function configurations, demonstrating the framework's ability to effectively cover target nets ranked by their cost function values.

\begin{table*}[!htbp]
\centering
\caption{Comparing the performance of \profuzz~with RFUZZ and DirectFuzz on various designs and highlighting the coverage improvement and speedup achieved by \profuzz~over DirectFuzz}
\label{tab:comparison}
\resizebox{\textwidth}{!}{%
\begin{tabular}{c|c|c|cc|cc|cccc}
\hline
\multirow{2}{*}{\begin{tabular}[c]{@{}c@{}}\textbf{Benchmark} \\\textbf{(Target Inst.)}\end{tabular}} &
\multirow{2}{*}{\begin{tabular}[c]{@{}c@{}}\textbf{\#Signals}\end{tabular}} &
\multirow{2}{*}{\begin{tabular}[c]{@{}c@{}}\textbf{Target Inst.}\\      \textbf{Cell \%}\end{tabular}} &
\multicolumn{2}{c|}{\textbf{\textcolor{black}{RFUZZ\cite{Laeufer'18}}}} &
\multicolumn{2}{c|}{\textbf{\textcolor{black}{DirectFuzz\cite{direct}}}} &
\multicolumn{4}{c}{\textbf{\profuzz} (This work)} \\ \cline{4-11} 
&
&
&
\multicolumn{1}{c|}{\textbf{Coverage}} &
\textbf{Time (s)} &
\multicolumn{1}{c|}{\textbf{Coverage}} &
\textbf{Time (s)} &
\multicolumn{1}{c|}{\textbf{Coverage}} &
\textbf{Time (s)} 
& \multicolumn{1}{|c|}{\textbf{Cov. Impv.\btrs}}
& \textbf{Speedup\bstars}\\ \hline
UART (Tx)
& 6
& 5.1\%
& 
\multicolumn{1}{c|}{100\%} & 7.35
& 
\multicolumn{1}{c|}{100\%} & 0.42
& 
\multicolumn{1}{c|}{100\%} & 1.31
&
\multicolumn{1}{|c|}{\textit{iso*}}
& 0.32
\\ \hline
UART (Rx)
& 9
& 6.9\%
&
\multicolumn{1}{c|}{88.89\%} & 4.95
&
\multicolumn{1}{c|}{88.89\%} & 1.71
&
\multicolumn{1}{c|}{100\%} & 1.43
&
\multicolumn{1}{|c|}{12.50\%}
& 1.20
\\ \hline
SPI
& 5
& 34.4\%
&
\multicolumn{1}{c|}{100\%} & 55.84
&
\multicolumn{1}{c|}{100\%} & 31.35
&
\multicolumn{1}{c|}{100\%} & 3.51
&
\multicolumn{1}{|c|}{\textit{iso*}}
& 8.93
\\ \hline
PWM
& 14
& 26.9\%
&
\multicolumn{1}{c|}{100\%} & 12.79
&
\multicolumn{1}{c|}{100\%} & 5.87
& 
\multicolumn{1}{c|}{100\%} & 2.76 
&
\multicolumn{1}{|c|}{\textit{iso*}}
& 2.13
\\ \hline
FFT
& 107
& 87\%
&
\multicolumn{1}{c|}{13\%} & 0.075
&
\multicolumn{1}{c|}{13\%} & 0.073
&
\multicolumn{1}{c|}{100\%} & 3.47
&
\multicolumn{1}{|c|}{669.23\%}
& 0.02 
\\ \hline
I2C
& 65
& 31\%
& 
\multicolumn{1}{c|}{98\%} & 13.73
&
\multicolumn{1}{c|}{98\%} & 8.49
&
\multicolumn{1}{c|}{100\%} & 4.57
&
\multicolumn{1}{|c|}{2.04\%}
& 1.86
\\ \hline
Sodar1Stage (CSR)
& 93
& 16.6\%
&
\multicolumn{1}{c|}{96.77\%} & 500.56
&
\multicolumn{1}{c|}{96.77\%} & 463.63
&
\multicolumn{1}{c|}{100\%} & 210.51
&
\multicolumn{1}{|c|}{3.34\%}
& 2.20
\\ \hline
Sodar1Stage (CtlPath)
& 68
& 0.3\%
&
\multicolumn{1}{c|}{100\%} & 694.42 
&
\multicolumn{1}{c|}{100\%} & 526.53
&
\multicolumn{1}{c|}{100\%} & 167.48
&
\multicolumn{1}{|c|}{\textit{iso*}}
& 3.14
\\ \hline
Sodar3Stage (CSR)
& 90
& 16.4\%
&
\multicolumn{1}{c|}{98.89\%} & 568.05
&
\multicolumn{1}{c|}{98.89\%} & 446.29
&
\multicolumn{1}{c|}{100\%} & 281.63
&
\multicolumn{1}{|c|}{1.12\%}
& 1.58
\\ \hline
Sodar3Stage (CtlPath)
& 66
& 0.3\%
&
\multicolumn{1}{c|}{100\%} & 1283.40
&
\multicolumn{1}{c|}{100\%} & 1034.86
&
\multicolumn{1}{c|}{100\%} & 236.94
&
\multicolumn{1}{|c|}{\textit{iso*}}
& 4.37
\\ \hline
\textbf{Average}
& 
& 
&
\multicolumn{1}{c|}{\textbf{89.56\%}} & \textbf{314.12}
&
\multicolumn{1}{c|}{\textbf{89.56\%}} & \textbf{251.92}
&
\multicolumn{1}{c|}{\textbf{100\%}} & \textbf{91.36}
&
\multicolumn{1}{|c|}{\textbf{11.66\%}}
& \textbf{2.76}
\\ \hline

\end{tabular}
}
\footnotesize
{\vspace{0.75mm}\btrs\textbf{Cov. Impv.}: Coverage Improvement in \% refers to the improvement achieved by \profuzz~over DirectFuzz. \hspace{1cm}\textbf{\textit{*iso}} denotes equivalent/same values.\\
\bstars\textbf{Speedup}: Refers to the reduction in execution time, calculated as the ratio of the execution time of DirectFuzz to that of \profuzz.}\raggedright
\end{table*}

\begin{figure}[!htbp]
  \centering
  \includegraphics[width=\columnwidth]{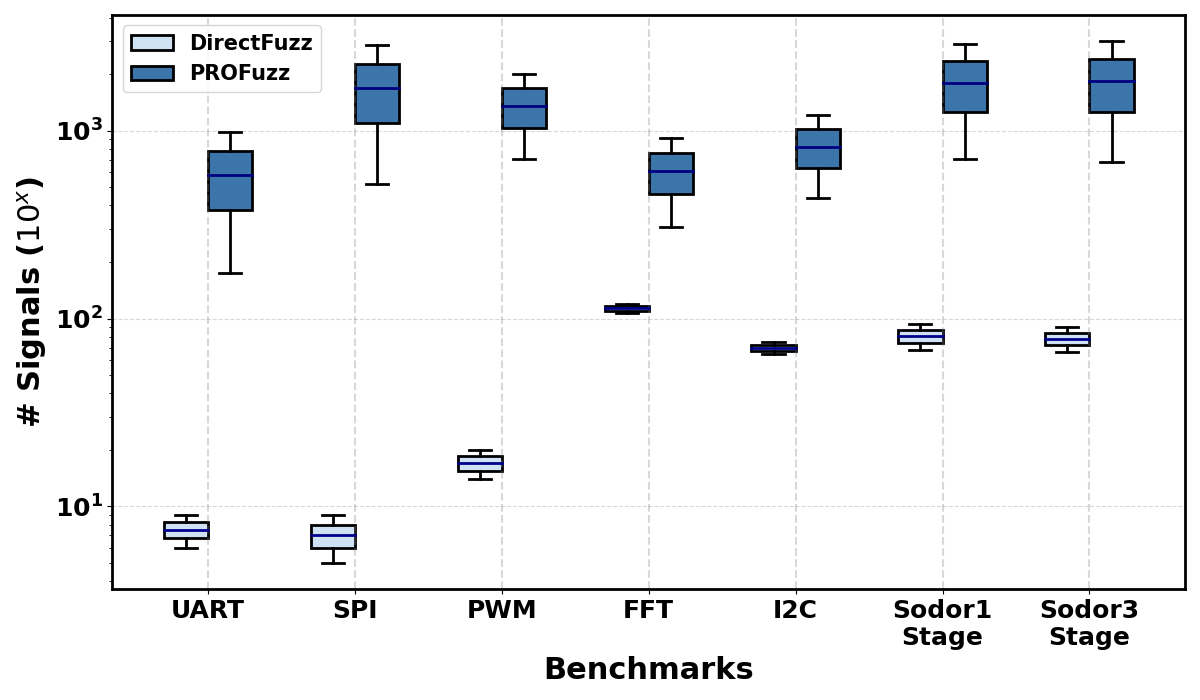} 
  \vspace{-2em}
  \caption{Whisker plots comparing the number of target sites handled by DirectFuzz and \profuzz.} 
  \label{fig:whisker}
\end{figure}

\subsection{Comparison with RFUZZ and DirectFuzz} \label{comparison}

Table \ref{tab:comparison} compares the performance of \profuzz~with existing state-of-the-art techniques, including RFUZZ\cite{Laeufer'18} and DirectFuzz\cite{direct}. As shown in previous subsections, \profuzz~demonstrates superior efficacy by covering a significantly larger set of target signals compared to RFUZZ and DirectFuzz. While covering fewer target sites in smaller designs (e.g., UART) is easier, \profuzz~performs better on more complex designs (e.g., Sodor5Stage). For a fair comparison, we used the same number of target signals as reported in DirectFuzz for each benchmark. Notably, RFUZZ and DirectFuzz struggle with \texttt{FFT}, where the number of target sites (107) is the highest among all tested designs. As seen in Table \ref{tab:comparison}, \profuzz~achieves higher average coverage, with 11.66\% improvement and a 2.76× speed-up over RFUZZ and DirectFuzz across designs of varying complexities.

\begin{figure*}[!htbp]
    \centering
    \subfloat[]{\includegraphics[width=0.5\textwidth]{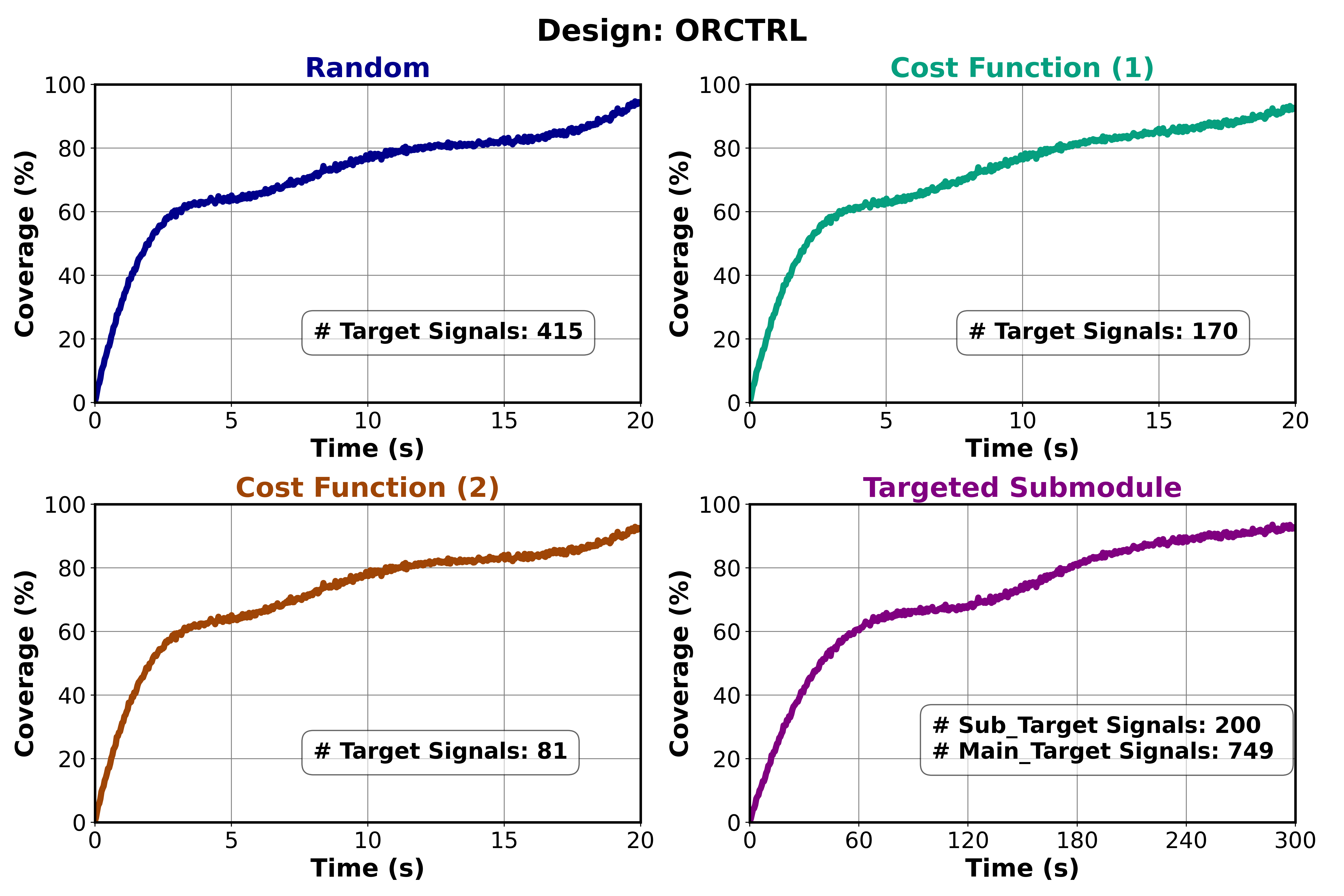}}
    \subfloat[]{\includegraphics[width=0.5\textwidth]{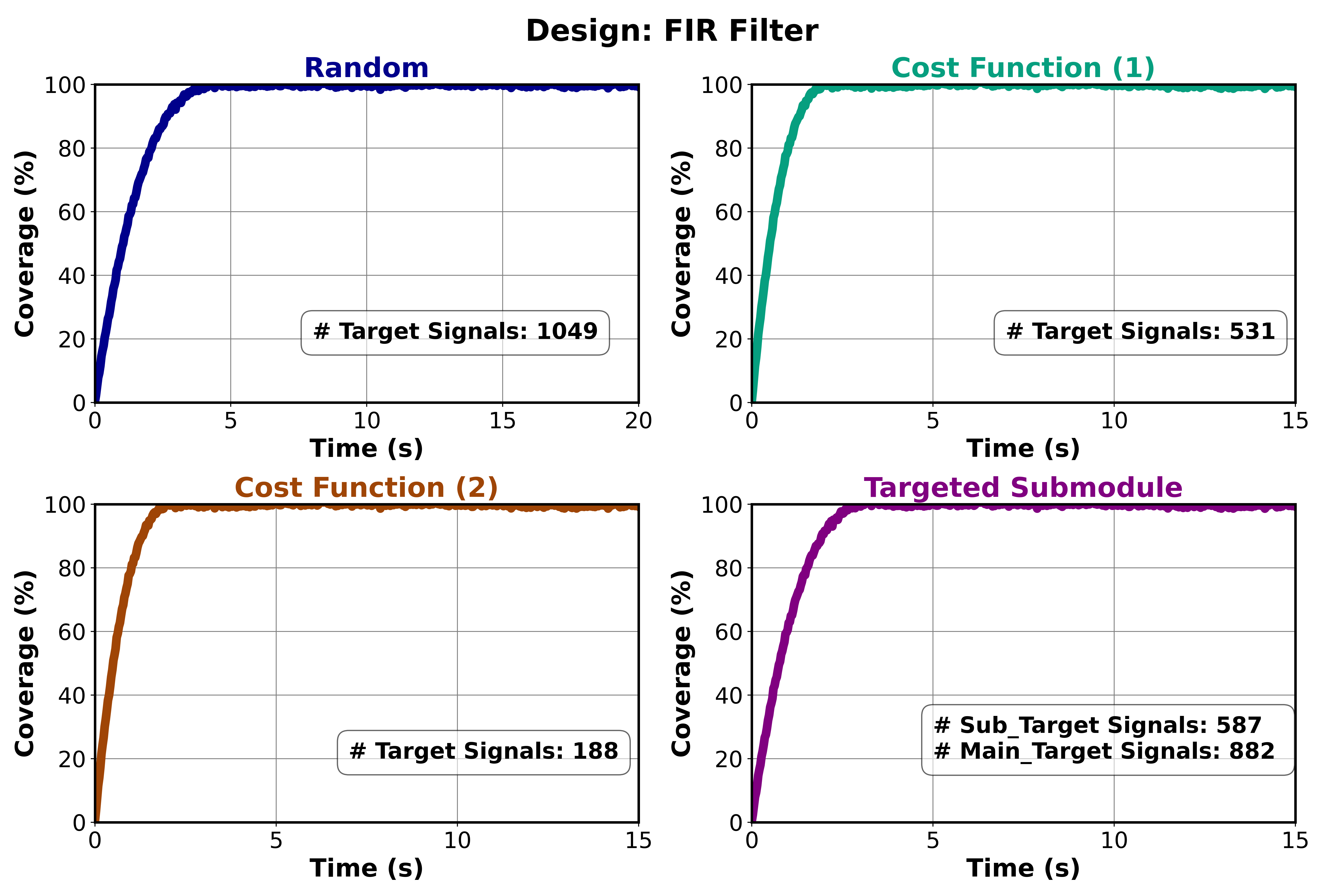}}
    \caption{Coverage convergence of \profuzz~with varying number of target sites selected based on various configurations (random, cost function, targeted submodule) for benchmarks: (a) ORCTRL and (b) FIR Filter.}
    \label{fig:performance}%
    \vspace{-1.5em}
\end{figure*}

Fig. \ref{fig:whisker} compares the number of target sites handled by \profuzz~and DirectFuzz across multiple benchmarks, showing a 2× to 30× increase in the number of target sites covered by \profuzz. This highlights the scalability and effectiveness of our approach, demonstrating its ability to handle a much larger number of target sites, particularly for complex hardware designs. As illustrated in Fig.~\ref{fig:performance}, \profuzz~demonstrates a steady improvement in target site coverage over time for all the configurations for \texttt{ORCTRL} and \texttt{FIR Filter}, highlighting the effectiveness of its directed fuzzing and efficient seed exploration in covering target sites.

\section{Conclusion} \label{conclusion}

This work introduces \profuzz, a Directed Graybox Fuzzing framework that enhances scalability by targeting specific regions within a hardware design. Unlike DirectFuzz, which struggles with large sets of target sites, \profuzz~supports signal-level activation across multiple modules and hierarchical regions, achieving 2×-30× more target sites covered with a 2.76× speedup over DirectFuzz. However, overall coverage depends on the nature of selected target sites, as some nets may hold constant logic values and cannot be toggled. Future research should focus on developing more intelligent, structure-aware mutation strategies specifically designed for hardware fuzzing, particularly for non-ISA-based inputs represented as raw bitstreams. Tailoring mutation engines to the hardware domain will be essential for enabling effective fuzzing across a broader range of complex hardware designs beyond traditional CPU cores or popular open-source projects.

\bibliographystyle{IEEEtran}
\bibliography{IEEEabrv,references}

\end{document}